\documentclass[namedreferences]{SolarPhysics}
\usepackage[optionalrh]{spr-sola-addons} 
\usepackage{graphicx}        
\usepackage{color}           
\usepackage{url}             
\usepackage{textcomp}

\newcommand{\etal}{{\it et al.}}
\newcommand{\eg}{{\it e.g.}}
\newcommand{\ie}{{\it i.e.}}
\newcommand{\cf}{{\it cf.}}

\renewcommand{\vec}[1]{ {\mathbf #1} }

\newcommand{\adv}{    {\it Adv. Spa. Res.}}

\newcommand{\araa}{   {\it Annu. Rev. Astron. Astrophys.}}
\newcommand{\aap}{    {\it Astron. Astrophys.}}

\newcommand{\aapr}{   {\it Astron. Astrophys. Rev.}}

\newcommand{\apj}{    {\it Astrophys. J.}}
\newcommand{\apjl}{   {\it Astrophys. J. Lett.}}

\newcommand{\mnras}{  {\it Mon. Not. Roy. Astron. Soc.}}
\newcommand{\nat}{    {\it Nature}}

\newcommand{\pasj}{   {\it Pub. Astron. Soc. Japan}}

\newcommand{\solphys}{{\it Solar Phys.}}

\newcommand{\ssr}{    {\it Space Sci. Rev.}}

\begin{document}

\begin{article}

\begin{opening}

\title{Imaging Observations of Quasi-Periodic Pulsatory Non-Thermal Emission in Ribbon Solar Flares \\ {\it Solar Physics}}

\author{I.V.~\surname{Zimovets}$^{1}$\sep
        A.B.~\surname{Struminsky}$^{1, 2}$
       }

\institute{$^{1}$ Space Research Institute, Russian Academy of Sciences, Profsoyuznaya str. 84/32,    Moscow, 117997 Russia\\
email: \url{ivanzim@iki.rssi.ru}\\
           $^{2}$ Institute of Terrestrial Magnetism, Ionosphere, and Radiowave Propagation, Russian Academy of Sciences, Troitsk, Moscow reg., 142190 Russia
              }

\begin{abstract}
Using RHESSI and some auxiliary observations we examine possible connections between spatial and temporal morphology of the sources of non-thermal hard X-ray (HXR) emission which revealed minute quasi-periodic pulsations (QPPs) during the two-ribbon flares on 2003 May 29 and 2005 January 19. Microwave emission also reveals the same quasi-periodicity. The sources of non-thermal HXR emission are situated mainly inside the footpoints of the flare arcade loops observed by the TRACE and SOHO instruments in the EUV range. At least one of the sources moves systematically both during the QPP-phase and after it in each flare that allows to examine the sources velocities and the energy release rate via the process of magnetic reconnection. The sources move predominantly parallel to the magnetic inversion line or the appropriate flare ribbon during the QPP-phase whereas the movement slightly changes to more perpendicular regime after the QPPs. Each QPP is emitted from its own position, which does not coincide with the origin of the previous pulsations. It is also seen that the velocity and the energy release rate don't correlate well with the flux of the HXR emission calculated from the sources. The sources of microwaves (observed by NoRH during the May 29 flare only) and thermal HXRs are situated near the apex of the loop arcade and are not stationary either. Almost all QPPs and some spikes of HXR emission during the post-QPP-phase reveal the soft-hard-soft spectral behavior indicating separate acts of electrons acceleration and injection, rather than modulation of emission flux by some kinds of magnetohydrodynamic (MHD) oscillations of coronal loops. In all likelihood, the flare scenarios based on the successively firing arcade loops are more preferable to interpret the observations, although we can not conclude exactly what mechanism forces these loops to flare up.       
\end{abstract}
\end{opening}

\section{Introduction}
     \label{S-Introduction}

Quasi-oscillatory phenomena are ubiquitous in nature and always attractive for researchers. This is also true for solar and stellar dynamic processes such as flares, filament oscillations (\eg, \opencite{Ballester06}), and magnetic loop oscillations (\eg, \opencite{Nakariakov05}; \opencite{Aschwanden06}). Sometimes, flux of non-thermal flaring emission reveals QPPs with periods in the range from milliseconds to minutes and in different frequencies from radio waves to HXRs (see review by \opencite{Aschwanden87}; and more recent \opencite{Aschwanden03}; \opencite{Nakariakov05}). It seems difficult to explain all variety of such QPPs by a unified physical process.

In this paper we concentrate only on long-periodic pulsations (with periods $>5$ \textrm{s} according to the classification given by \opencite{Aschwanden03}) simultaneously observed in the HXR and microwave ranges which can be manifestations of single population of accelerated electrons in the solar atmosphere, because of similarity of these emissions' light curves during many flares (\eg{}, see for details a review by \opencite{Bastian98}). It's accepted that a bulk of non-thermal HXRs in solar flares is the thick-target bremsstrahlung radiation of non-thermal electrons ($20-100$ \textrm{keV}) precipitating along foots of magnetic loops into the chromosphere, whereas the microwave emission is produced via the gyrosynchrotron process of accelerated electrons ($>100$ \textrm{keV}) interacting with magnetic field. QPPs with such periods often assumed to be generated by some kind of MHD oscillations of a flare coronal loop with fixed footpoints which contains already accelerated electrons (apparently, the idea have appeared firstly in \opencite{Rosenberg70}, although it was used for interpretation of the fast $\sim 1$ \textrm{s} broad band fluctuations). Such MHD oscillations can cause periodic oscillations of the magnetic loop cross section. Consequently, the magnetic field strength and magnetic mirroring can be modulated and the flux of already accelerated electrons, precipitating towards the loop's footpoints, becomes modulated too, causing QPPs of the HXR emission (\eg{}, \opencite{Zaitsev82}; \opencite{Zaitsev89}). QPPs of the microwave emission can naturally appear in this scenario, because of magnetic field variations. \inlinecite{Brown75} developed a bit different model of the QPPs based on a betatron action of the flaring vibrating magnetic bottle on the already accelerated and trapped electrons changing their spectrum and flux.      

Another class of models based on the idea, that MHD oscillations of coronal loops can modulate efficiency of accelerating process. Thus, \inlinecite{Asai01} suggested under the \inlinecite{Tsuneta98} flare model, that QPPs can be caused by quasi-periodic modulation of the fast shock's length (that can accelerate electrons by the first-order Fermi process) via oscillations of an underlying flaring loop. Be guided by their own observations of the large transequatorial loop, which places near the flaring region of thermal X-ray emission with QPPs, \inlinecite{Foullon05} suggested another scenario according to which QPPs can be caused by periodic pumping of electrons into an acceleration region by the linked oscillating loop. \inlinecite{Nakariakov06} proposed the similar model but with modifications, according to which a modulation of electric current density in an acceleration region can be produced by penetrating fast magnetoacoustic oscillations from a non-flaring oscillating loop. In its turn oscillations of current density can produce oscillations of the magnetic reconnection process via anomalous resistivity and, as a result, an oscillatory acceleration of electrons. Unfortunately, to the best of our knowledge, there were not yet simultaneous observations of EUV loop oscillations and QPPs of microwave and non-thermal HXR emission, although spatially resolved oscillations of a flare loop in the microwave range were found \cite{Nakariakov03}. This remains possibilities for QPPs to be originated not only by a non-reconnecting oscillatory coronal loop with fixed footpoints.    

According to the standard model of erupting solar flares (the CSHKP model; \opencite{Carmichael64}; \opencite{Sturrock66}; \opencite{Hirayama74}; \opencite{Kopp76}), an emerging filament can stretch overlying magnetic field lines and form a quasi-vertical current sheet, where magnetic field lines can reconnect releasing the stored magnetic energy that can be partially converted into accelerated charged particles and consequently in the HXR and microwave emission. The released energy rate, $dW/dt$, can be estimated as a product of the area of the reconnection region and the Poynting flux, $\left|\vec{S}\right| \sim \left|\vec{E_{cor}}\times\vec{B_{cor}}\right|$ (\opencite{Isobe02}), if one assume that all energy is released (this is a serious assumption). Coronal electric field, $E_{cor}$, can be estimated as $v_{in}B_{cor}$, where $v_{in}$ is the inflow velocity of coronal magnetic field lines with $B_{cor}$ drawing into the reconnection region. Thus,
\begin{equation}  \label{Eq_1}
     dW/dt \sim v_{in}B_{cor}^{2},
   \end{equation}
if one assumes the constant area of the reconnection region.

The inflow velocity can be estimated if one observes an upward motion of flaring loops with a high cadence and spatial resolution (say, in EUV or in soft X-rays), that can be successively involved in the moving reconnection site. This process reveals itself via EUV or H$\alpha$ flare ribbons separation preferably perpendicular to the magnetic inversion line and via growing loop arcades. \inlinecite{Sui04} using \textit{Reuven Ramaty High-Energy Solar Spectroscopic Imager} (RHESSI; \opencite{Lin02}) found correlation between the rate of the flare loops upward motion and HXR flux. Theoretically, motions of the reconnection site can have an oscillatory behavior, as it was shown by numerical MHD modeling \cite{Chen99}. But one exciting observation with the \textit{Transition Region and Coronal Explorer} (TRACE; \opencite{Handy99}) of the oscillatory shrinkage of flaring loops in 195 \textrm{\r{A}} \cite{Li06} did not reveal clearly an oscillatory flux of the RHESSI HXR emission, as it could be expected. Nevertheless, theoretical evidences of an oscillatory regime of the magnetic reconnection under the coronal conditions were shown in quantity (\eg{}, \opencite{Tajima87}; \opencite{Kliem00}; \opencite{Ofman06}) and we have to wait more precise observations to examine this possibility. 
 
Another possibility of a QPP-generating mechanism consists in successive acts of energy release in different places of flaring arcade (\eg, \opencite{Vorpahl76}). As multiple and quite random non-periodic pulsations of non-thermal emission are often observed during ribbon flares, but QPPs are rare phenomena, QPPs may be just a special case of non-periodic pulsations under a unique configuration of magnetic field. An impressive example of pulsatory RHESSI HXR emission (but without obvious periodicity) from a two-ribbon flare was presented by \inlinecite{Grigis05}, were analysing motions of footpoint-like HXR sources. Authors showed that each observed ``elementary flare bursts'' were not modulated by a single oscillating flare loop with stationary footpoints or by an oscillating reconnection process in a confined flare region, rather by a process of reconnection site motion along the flare arcade. Since detailed observations of coronal source motions are succeeded rarely, but motions of HXR footpoints are observed in quantity (using Yohkoh and RHESSI), $v_{in}$ was related to a component of velocity of footpoint-like HXR sources which is perpendicular to the magnetic inversion line, $v_{ftp}^{\bot}$, in the frame of the simple 2D reconnection model (under the assumption of the translation simmetry along a flare arcade; see review by \opencite{Priest02}). Three extra assumptions were made for this: (a) just reconnected field lines have footpoints close to the footpoints of previous lines; (b) the magnetic flux conservation is satisfied, \ie{}, $v_{in}B_{cor}=v_{ftp}^{\bot}B_{ftp}$, where $B_{ftp}$ is the magnetic field in the footpoints of a reconnecting coronal loop; (c) magnetic field in the recconection site in the corona is proportional to the magnetic field in the footpoints of the loops. Thus, the relation~(\ref{Eq_1}) can be rewritten as
\begin{equation}  \label{Eq_2}
     dW/dt \sim v_{ftp}^{\bot}B_{ftp}^{2}.
   \end{equation}
\inlinecite{Grigis05} did not find correlation between the rate of footpoints motion perpendicular to flare ribbons and the HXR flux as it can be expected from the simple 2D models. Authors suggested an interesting idea amending these models: a filament may erupt unevenly along its length.    

Several flares accompanied by motions of the footpoint-like HXR sources along the magnetic inversion line or flare ribbons were really found in quantity (\eg{}, \opencite{Fletcher02}; \opencite{Bogachev05}; \opencite{Jing07}; \opencite{Gan08}) using HXR data from Yohkoh and RHESSI. Motions of H$\alpha$ bright kernels along the flare ribbons were also found, \eg{}, by \inlinecite{Qiu02}. Authors of the papers also concluded that it was difficult to interpret their observations in terms of the simple 2D reconnection model. Most probably, the source of primary energy release can somehow propagate mainly along the sheared flaring arcade in some ribbon flares. Quasi-pulsatory behavior of non-thermal emission may be a consequence of sequentially bursting adjacent arcade loops with quite similar physical properties (size, magnetic field, density). 

One can see that the problem of solar flare QPPs in non-thermal range is quite intricate and is still far from its complete solving. Many possibilities were supposed to interpret  quasi-periodicities of the observed emission light curves. Recently, \inlinecite{Li08} analysing one solar flare event accompanied by QPPs of HXR and microwave emission came to the conclusion that their event ``cannot be easily explained with the existing mechanisms''. Thus, it seems reasonable to find and analyse another flares with QPPs of non-thermal emission using imaging observations.

The main goal of the paper is to examine connections between spatial and temporal behavior of non-thermal HXR emission observed by RHESSI during a quite rare phenomenon - the clearly marked phase of minute QPPs during the solar flares on 2003 May 29 and 2005 January 19, and to verify if some changes occur in morphology of the HXR sources under transition from the QPP-phase to the post-QPP-phase. Unfortunately, it is impossible to examine dynamics of coronal magnetic structures in the EUV range, because of the instrumental problems. One extra goal is to examine validity of the simple 2D reconnection model for the concerned events using the relation~(\ref{Eq_2}). RHESSI observations of the January 19 flare, but without a detailed analysis of spatial evolution of the HXR sources, was reported by \inlinecite{Nakariakov06} and \inlinecite{Ofman06} as an impressive example of the flare with QPPs. Authors proposed different mechanisms to generate the observed QPPs. \inlinecite{Saldanha08} and \inlinecite{Grigis08} reported the detailed HXR spectral analysis and also some imaging analysis of the same event but without giving consideration to the QPP-problem. Detailed spectral analysis of non-thermal emission during the QPP-phase of the 2003 May 29 flare was reported by \inlinecite{Minoshima08} and by \inlinecite{Ning07}. Thus, we will restrict ourself only by the detailed imaging analysis of these QPP-events.
               
\section{Flare of 2003 May 29} 
 \label{S-May}

\subsection{Features of the flare} 
  \label{S-MayFeature}  
\begin{figure}    
   \centerline{\includegraphics[width=0.99\textwidth,bb=136 283 451 563,clip=]{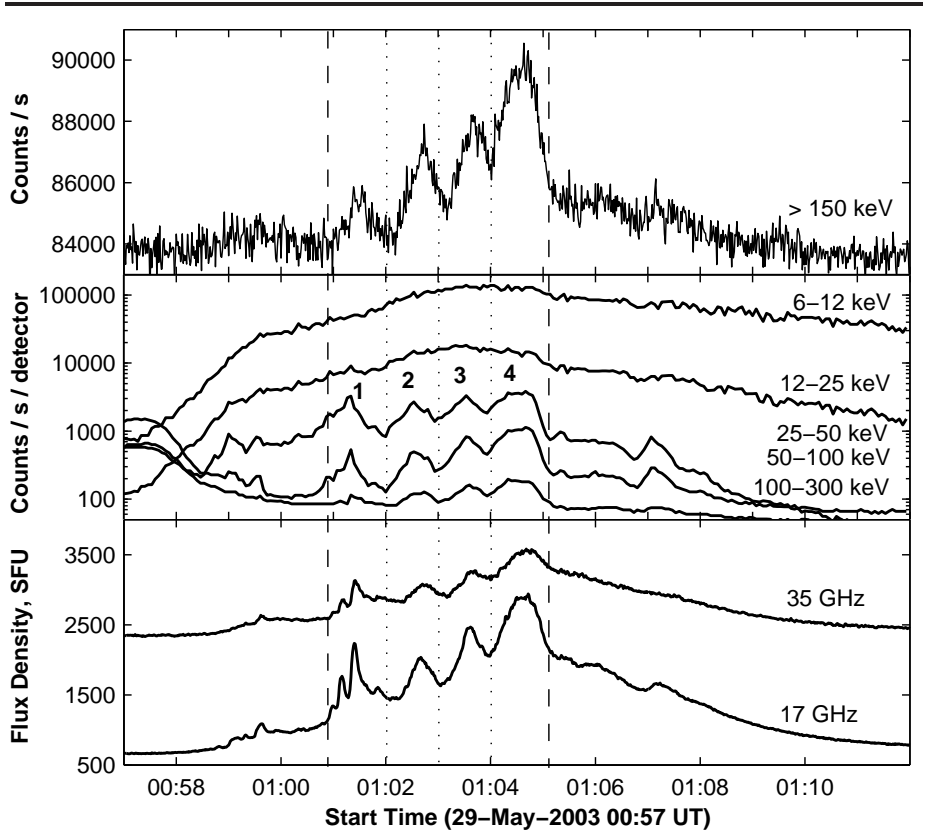}
              }
              \caption{\emph{Top panel:} ACS count rate of $>150$ \textrm{keV} HXR emission accumulated over 1 s, background level is not subtracted. \emph{Middle panel:} RHESSI corrected count rates at five energy bands (counts at $50-100$ and $100-300$ \textrm{keV} bands are multiplied by 1.1 and 0.7 respectively for clarity) accumulated by the detectors 1, 3, 4, 5, 8, 9. The time resolution is 4 \textrm{s}. \emph{Bottom panel:} 1-s resolution time profiles of total NoRP flux densities at 17 and 35 \textrm{GHz}, background level is not subtracted. The QPP-phase is between two vertical dashed lines. The vertical dotted lines and bold numerals indicate the QPPs. 
                      }   \label{F-1}
   \end{figure}

A two-ribbon solar flare of GOES soft X-ray importance X1.2 with the heliographic coordinates (S06\textdegree{}, W37\textdegree{}) from the NOAA Active Region 10365 started at 00:51 UT, had its peak and its end at 01:05 UT and 01:12 UT respectively (\textit{Solar-Geophysical Data}). HXR emission from this flare was clearly observed by RHESSI mainly in the $<300$ \textrm{keV} energy range; by the \textit{Anti-Coincidence System} (ACS) of the SPI spectrometer aboard INTEGRAL (\eg, \opencite{Vedrenne03}, and references therein) in the available energy range $>150$ \textrm{keV}; and also by the SONG device onboard the ``CORONAS-F'' satellite within two energy channels: $60-150$ and $150-500$ \textrm{keV} (Miagkova, private communication, 2008). Quite large RHESSI parasitic counts of non-solar origin were observed from the end of the spacecraft's night-time at about 00:42 UT till 01:00 UT and hampered observing the first flare's nonthermal bump started at about 00:57--00:58 UT in the microwave range according to observations of Nobeyama Radio Polarimeters (NoRP; \opencite{Nakajima85}) and in the HXR range observed by ACS and SONG (Figure~\ref{F-1}). Note, that the fine structure of the ACS light curve is mainly an instrumental noise, rather than a fine structure of the solar flare emission. 

At the least four growing QPPs of HXR and microwave emission with about one-minute duration are clearly seen on Figure~\ref{F-1} between 01:01 and 01:05 UT. We denote this time period as a QPP-phase. The first and the last of these pulsations are clearly composed of several shorter spikes. Apparently, one ulterior spiky structure was between 00:58 and 01:01 UT and another one was between 01:05 and 01:08 UT. These structures were composed of several shorter spikes having smaller modulation depths. Time profiles of the RHESSI thermal HXR emission have smoother characters and don't reveal QPPs (Figure~\ref{F-1}).

According to spectral analysis of HXR and microwave emission during the QPP-phase of this flare performed by \inlinecite{Minoshima08} spectrum of the HXR emission in 40 - 250 \textrm{keV} range at each time is fitted well by a double-power law function and microwave emission above 17 \textrm{GHz} is the optically thin non-thermal gyrosynchrotron emission. \inlinecite{Minoshima08} also found that temporal behavior of spectral index of the non-thermal gyrosynchrotron emission is similar to that of high-energy component of the HXR emission indicating that these emissions are due to the same populations of accelerated electrons. Moreover the time profile of the spectral index of low-energy non-thermal HXR component reveals the ``soft-hard-soft'' behavior for each QPPs except the last one. This spectral behavior is considered to be an intrinsic signature of an elementary act of electron acceleration in flares (\eg, \opencite{Grigis04}; \opencite{Battaglia06}). Thus, we may suppose, that the observed QPPs of HXR and microwave emission may be due to individual acts of electrons acceleration and injections into coronal loops, rather than due to modulation of electrons trapping and precipitation after a single act of their energization (see Section~\ref{S-Discussion} for discussions).          

\subsection{Images of the flare region} 
  \label{S-images1}
  
To examine spatial evolution of thermal and non-thermal X-ray sources during the QPP-phase between 01:01 and 01:05 UT and after it, till 01:08 UT, when the HXR sources were still quite clearly visible, we implement Clean and Pixon algorithms to the RHESSI data with the best FWHM resolution of 3.9$2^{\prime\prime}$ obtained by detectors $2-8$ \cite{Hurford02}. Series of images integrated over 8 and 16 \textrm{s} (about 2 and 4 RHESSI periods of rotation) are obtained in the 12 - 18 \textrm{keV} (thermal source) and in the 30 - 100 \textrm{keV} (non-thermal source) energy ranges with pixel size of $1^{\prime\prime}$. The division into these energy bands is made according to spectral analysis of the RHESSI data implemented using the OSPEX code which is considered an object-oriented development of the SPEX (\opencite{Smith02}, and references therein). Two changes of state of the RHESSI attenuators were made during the time period considered: the first one was at 01:05:44 UT (A3\textrightarrow{A1}) and the second one was at 01:06:12 UT (A1\textrightarrow{A3}), that made a gap in the sequences of the images. To partially fulfill the gap we make several additional images with different durations for this time interval.   

\begin{figure}    
   \centerline{\hspace*{0.01\textwidth}
               \includegraphics[width=0.494\textwidth,bb=83 27 389 323,clip=]{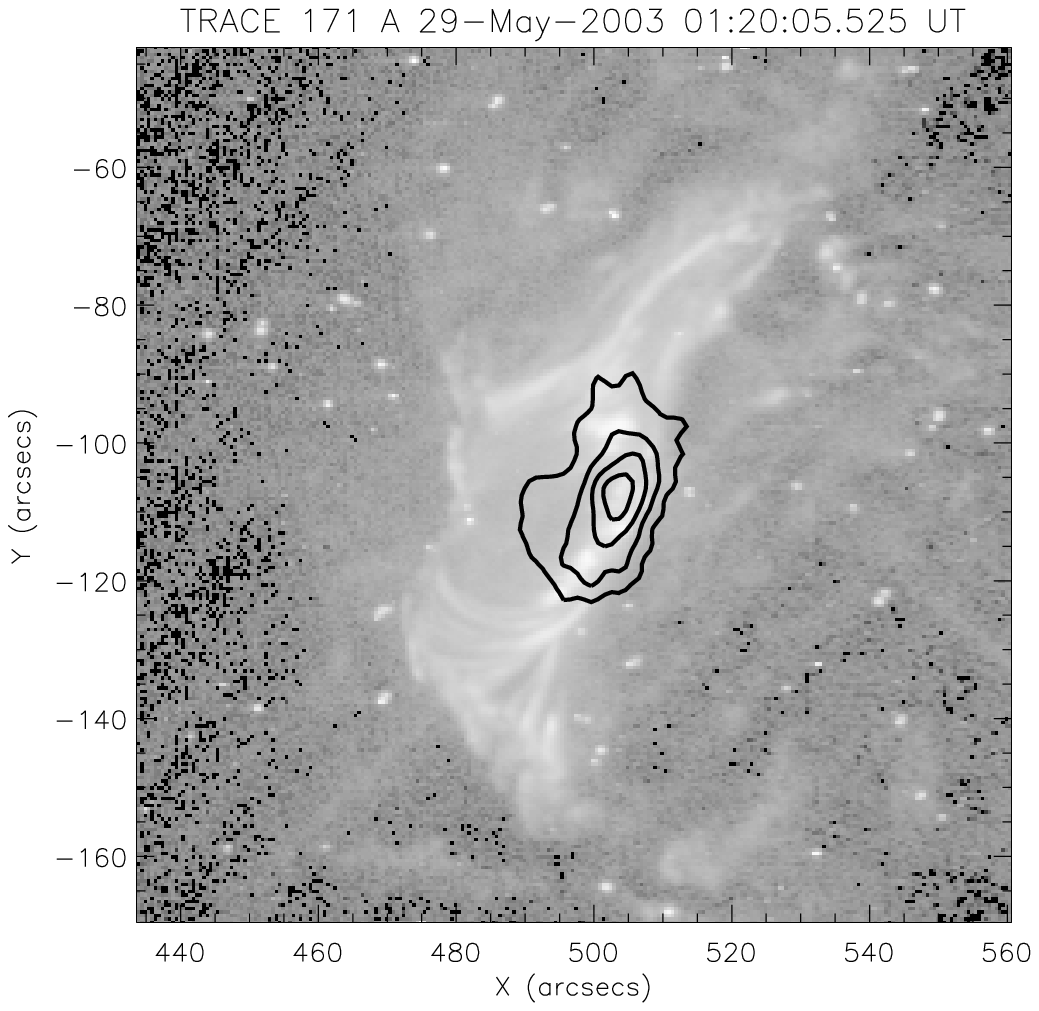}
               \includegraphics[width=0.494\textwidth,bb=83 27 389 323,clip=]{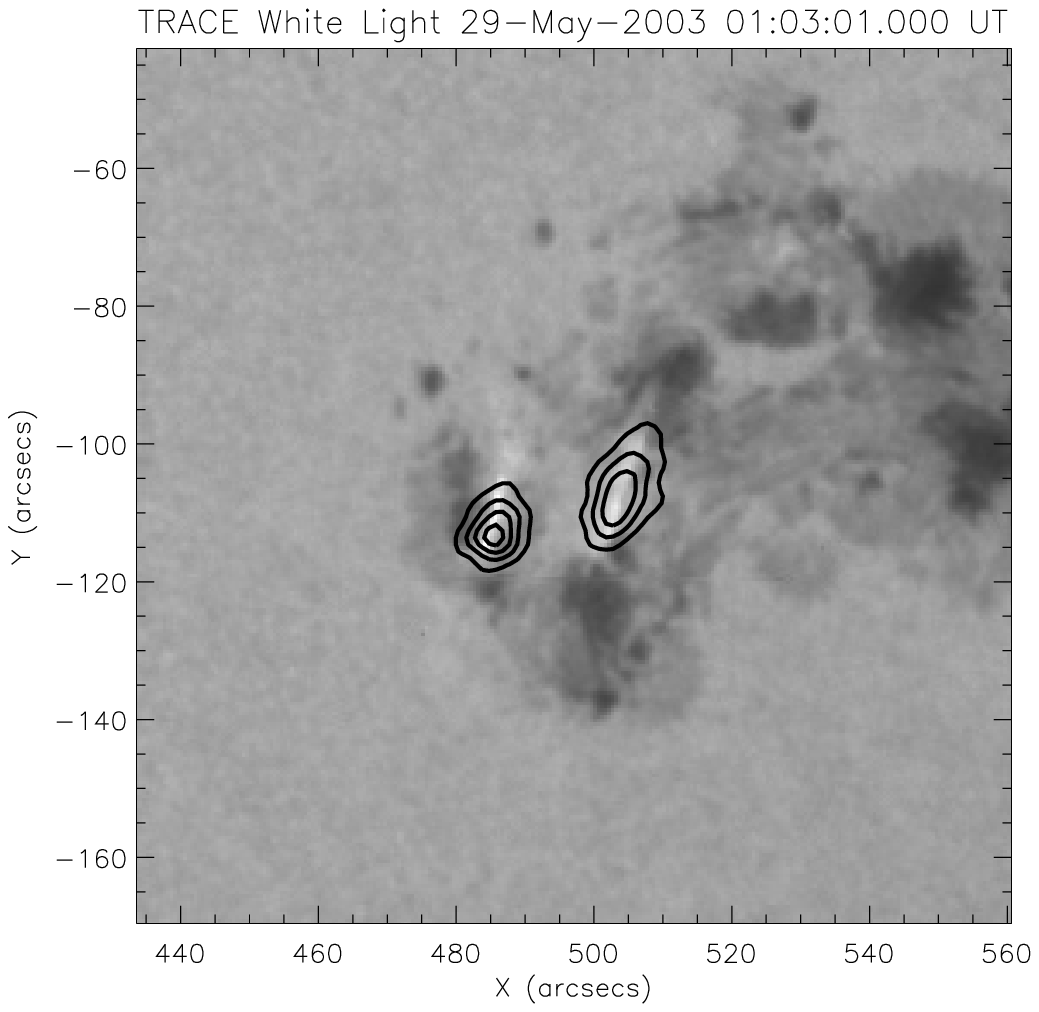}
              }
     \vspace{-0.43\textwidth}   
     \centerline{\Large \bf     
      \hspace{0.1 \textwidth}  \color{white}{(a)}
      \hspace{0.42\textwidth}  \color{white}{(b)}
         \hfill}
     \vspace{0.4\textwidth}    
   \centerline{\hspace*{0.01\textwidth}
               \includegraphics[width=0.494\textwidth,bb=83 27 389 323,clip=]{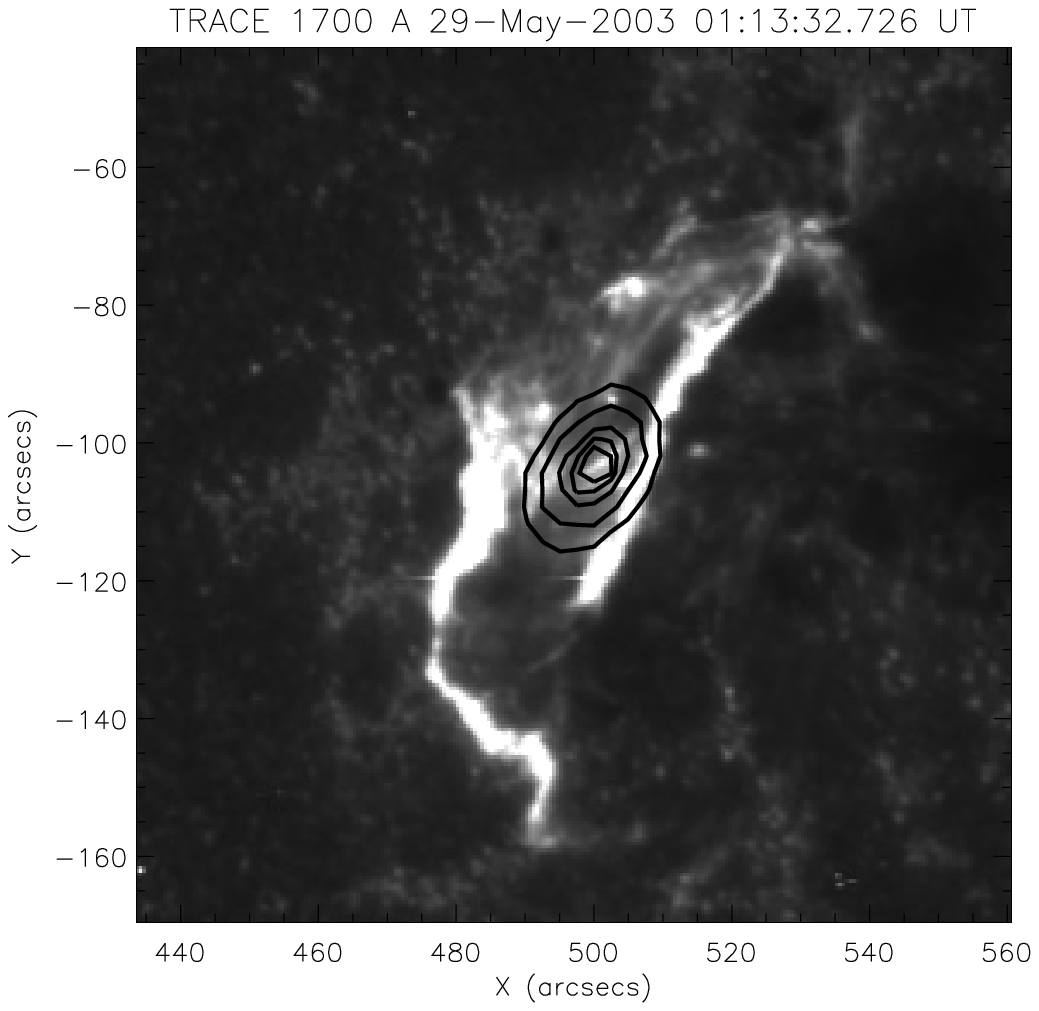}
               \includegraphics[width=0.494\textwidth,bb=83 27 389 323,clip=]{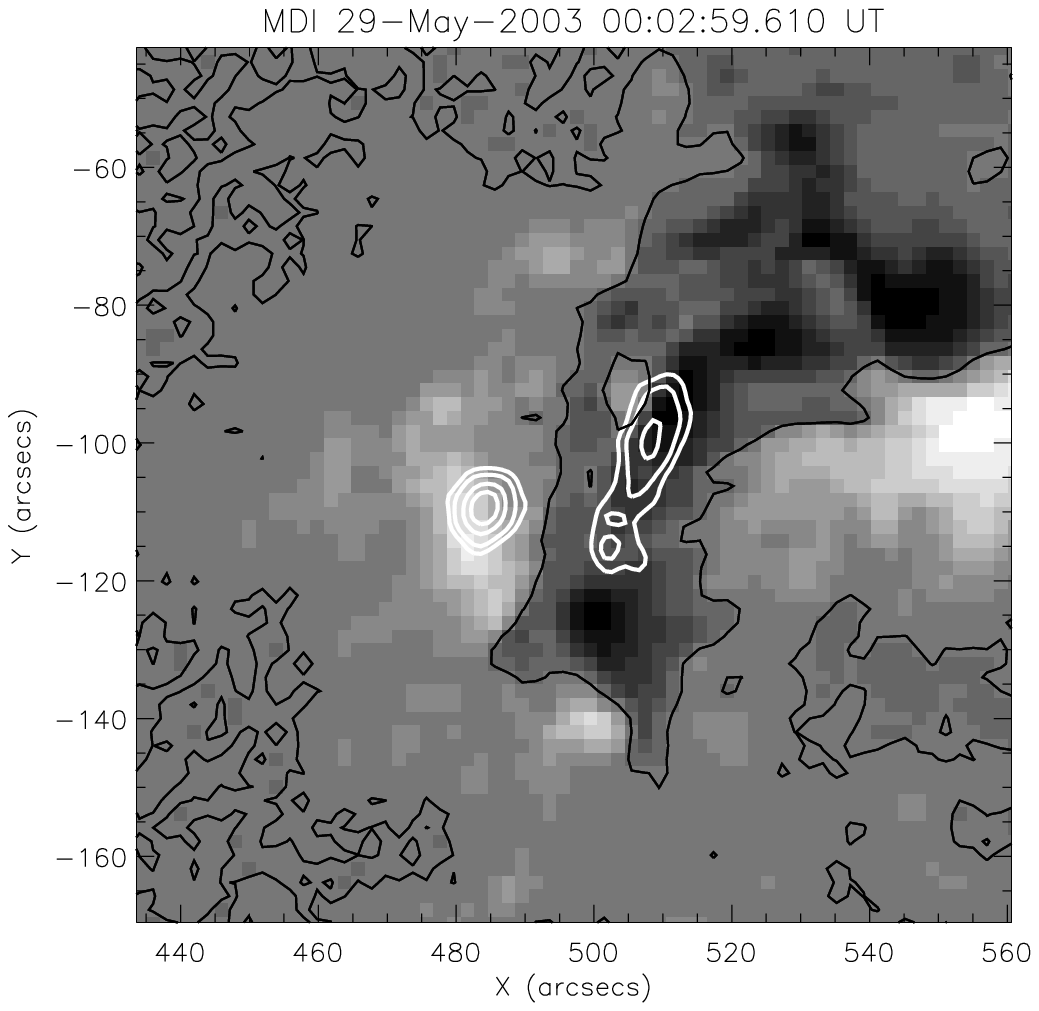}
              }
     \vspace{-0.43\textwidth}   
     \centerline{\Large \bf     
      \hspace{0.1 \textwidth} \color{white}{(c)}
      \hspace{0.42\textwidth}  \color{white}{(d)}
         \hfill}
     \vspace{0.41\textwidth}    
\caption{Typical location of emission in the region of the 2003 May 29 two-ribbon solar flare. \textbf{(a)} The TRACE 171 \textrm{\r{A}} image overlaid by the RHESSI $12 - 18$ \textrm{keV} contours (30\%, 50\%, 70\%, 90\%, black lines) made at 01:04:04 - 01:04:20 UT. \textbf{(b)} The TRACE white light image overlaid by the RHESSI $30-100$ \textrm{keV} contours (30\%, 50\%, 70\%, 90\%, black lines) made at 01:02:40 - 01:02:56 UT. \textbf{(c)} The TRACE 1700 \textrm{\r{A}} image overlaid by the contours (40\%, 60\%, 80\%, 90\%, 95\%, black lines) of the NoRH 17 \textrm{GHz} brightness temperature  made at 01:01:29 UT. \textbf{(d)} The SOHO/MDI magnetogram overlaid by the RHESSI $30-100$ \textrm{keV} contours (30\%, 40\%, 60\%, 80\%, white lines) made at 01:04:20 - 01:04:36 UT. Black lines correspond to the magnetic inversion lines. All images were rotated to the RHESSI $12 - 18$ \textrm{keV} one.
                      }              
   \label{F-2}
   \end{figure}

Figure~\ref{F-2} shows typical morphology of emission in the flare region. The TRACE 171 \textrm{\r{A}} image (Figure~\ref{F-2}(a)), made at about 01:20:06 UT after the flare's end, indicates a post-flare arcade of magnetic loops or an ``eruptive flare loop system'' as it was sounded by \inlinecite{Svestka07}. The flare was apparently an eruptive event. Although no manifestation of an active H$\alpha$ filament was found (\textit{Solar-Geophysical Data}), the fast halo coronal mass ejection was observed by the telescope 2 of the Large Angle and Spectrometric Coronagraph onboard \textit{Solar and Heliospheric Observatory} satellite (SOHO/Lasco/C2) firstly at 01:27:12 UT (\textit{Solar-Geophysical Data}).

The RHESSI $12-18$ \textrm{keV} contours place mainly inside the brightest region of the TRACE 171 \textrm{\r{A}} arcade and nearly coincide with the location of the source of the NoRP 17 \textrm{GHz} emission between two flare ribbons (Figure~\ref{F-2}(c)), which are clearly seen in the TRACE 1700 \textrm{\r{A}} image  made at about 01:13:33 UT (note, that the location of the microwave and thermal HXR emission seems to be closer to the eastern ribbon rather than to the apparent arcade's apex, probably because of the projection effect). Using the NoRH observations \inlinecite{Minoshima08} showed that the microwave source placed inside the region of the reduced degree of polarization. These facts may indicate that microwaves were mainly emitted from the arcade apex. Figure~\ref{F-2}(b, d) shows that the sources of RHESSI $30-100$ \textrm{keV} emission place mainly inside the eastern sunspot of positive magnetic polarity and between two western sunspots of negative polarity. These places coincide well with the footpoints of the arcade and with the flare ribbons (\cf{} with Figure~\ref{F-2}(a, c)). All images in Figure~\ref{F-2} were rotated to the RHESSI $12-18$ \textrm{keV} image made at 01:04:04 - 01:04:20 UT using a procedure \verb+drot_map+ written by Dominic Zarro within the SolarSoftWare.  

\begin{figure}    
   \centerline{\includegraphics[width=0.99\textwidth, bb=126 282 461 565,clip=]{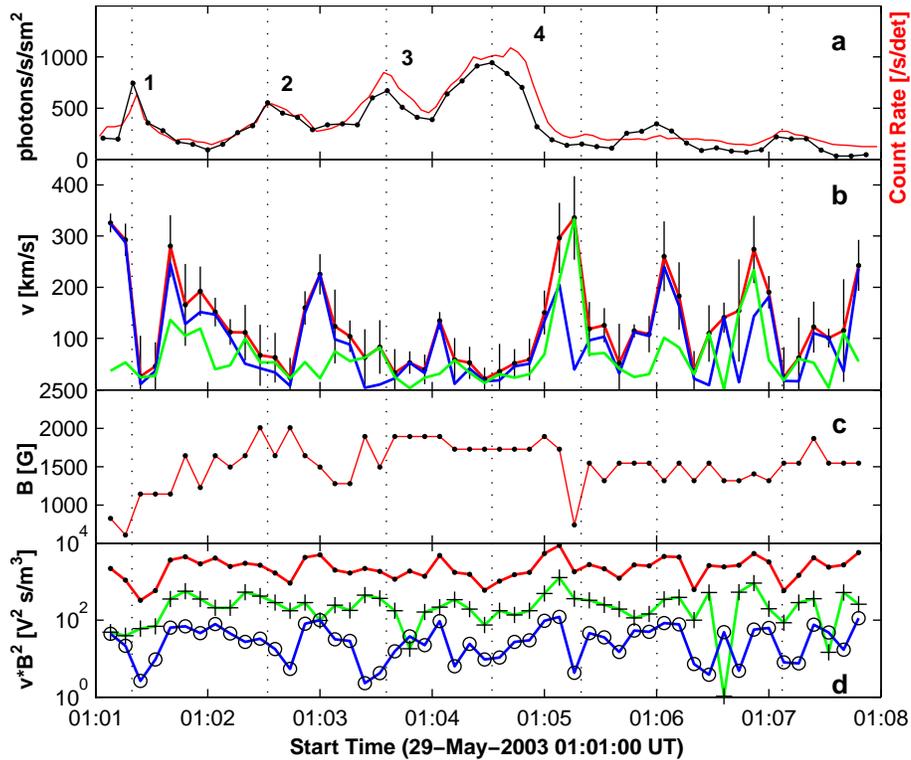}
              }
              \caption{Time profiles of calculated characteristics of the eastern non-thermal HXR source of the 2003 May 29 solar flare. \textbf{(a)} The RHESSI full-sun corrected count rate at $50-100$ \textrm{keV} accumulated over Ge detectors 1, 3, 4, 5, 8, 9 (red line). Time resolution is 4 \textrm{s}. HXR flux calculated from the Pixon images of the eastern source at $30-100$ \textrm{keV} (black dotted line; flux is multiplied by 10). Time of the images accumulation is 8 \textrm{s}. \textbf{(b)} Full velocity of the eastern source (red line). Errors are marked by vertical black lines. Absolute values of the full velocity components parallel and perpendicular to the averaged magnetic inversion line are marked by blue and green lines respectively. \textbf{(c)} The SOHO/MDI line-of-sight photospheric magnetic field below the eastern HXR source (red line with black dots). \textbf{(d)} $v_{ftp}B^{2}_{ftp}$ (red line with black dots), $v^{\|}_{ftp}B^2_{ftp}$ (divided by 50; blue line with circles) and $v^{\bot}_{ftp}B^{2}_{ftp}$ (divided by 5; green line with black crosses). 
                      }   \label{F-3}
   \end{figure}

The eastern HXR source in $30-100$ \textrm{keV} range is on average more compact and brighter than the western one, which reveals a gaunt shape along the western ribbon in some moments during the flare (Figure~\ref{F-2}(d)). Time profile of the HXR flux at $30-100$ \textrm{keV} from the eastern source correlates well with the full sun RHESSI corrected count rate at $50-100$ \textrm{keV} (Figure~\ref{F-3}(a)), indicating a direct link of the eastern source with the primary origin of the QPPs. To calculate the time profile of the HXR flux from the source it was encircled with the circuit of $R\approx6^{\prime\prime}$ in each images and the flux of photons from the obtained area is calculated in the frame of the RHESSI branch of the SolarSoftWare.

We mainly pay our attention to the eastern HXR source because of its simplicity, systematic motions and its links to the QPPs. Some words about the western HXR sources are below. We also must emphasize that the brightest regions of the thermal HXR emission are situated in different places of the arcade apex in the course of the flare development that can be interpreted as motion of the energy release site. But detailed analysis of this source motion is not available because of its complexity. 

\begin{figure}    
   \centerline{\hspace*{0.01\textwidth}
               \includegraphics[width=0.49\textwidth,bb=148 278 428 575,clip=]{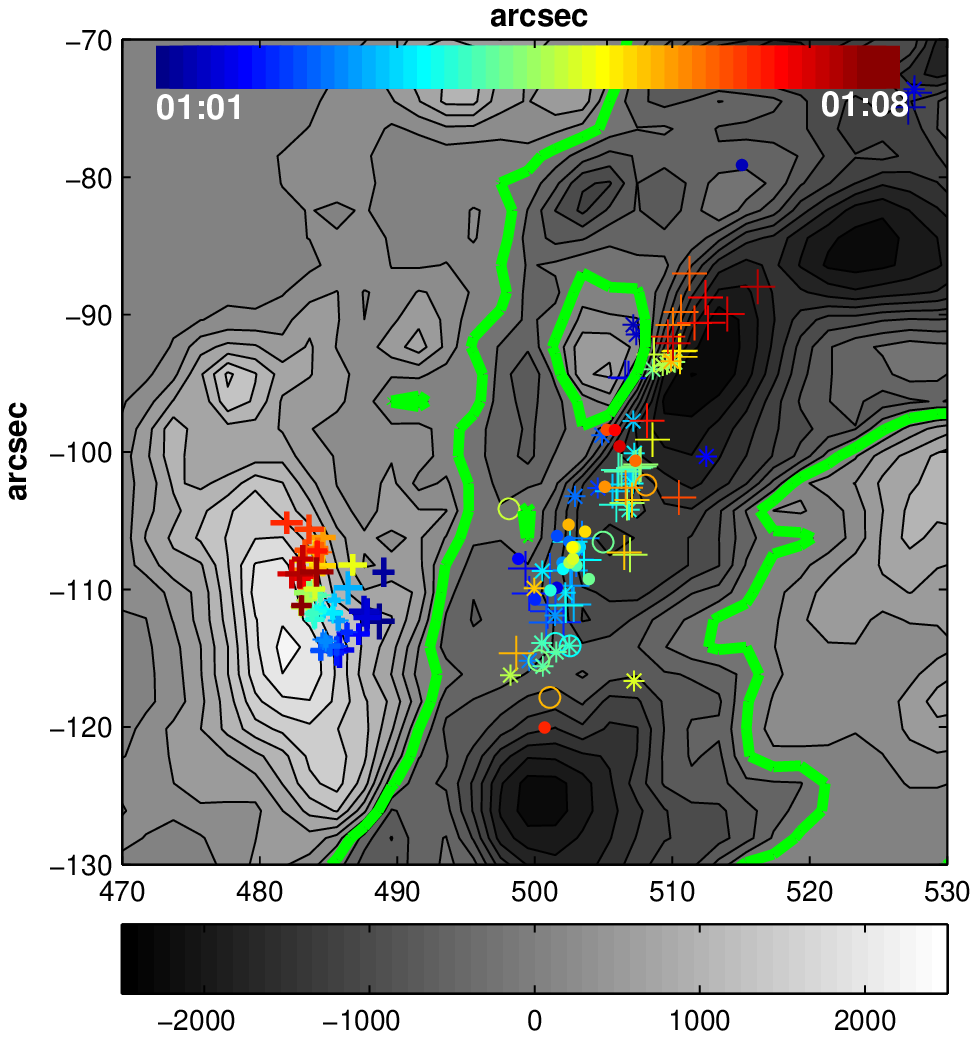}
               \hspace*{-0.01\textwidth}
               \includegraphics[width=0.49\textwidth,bb=188 296 444 569,clip=]{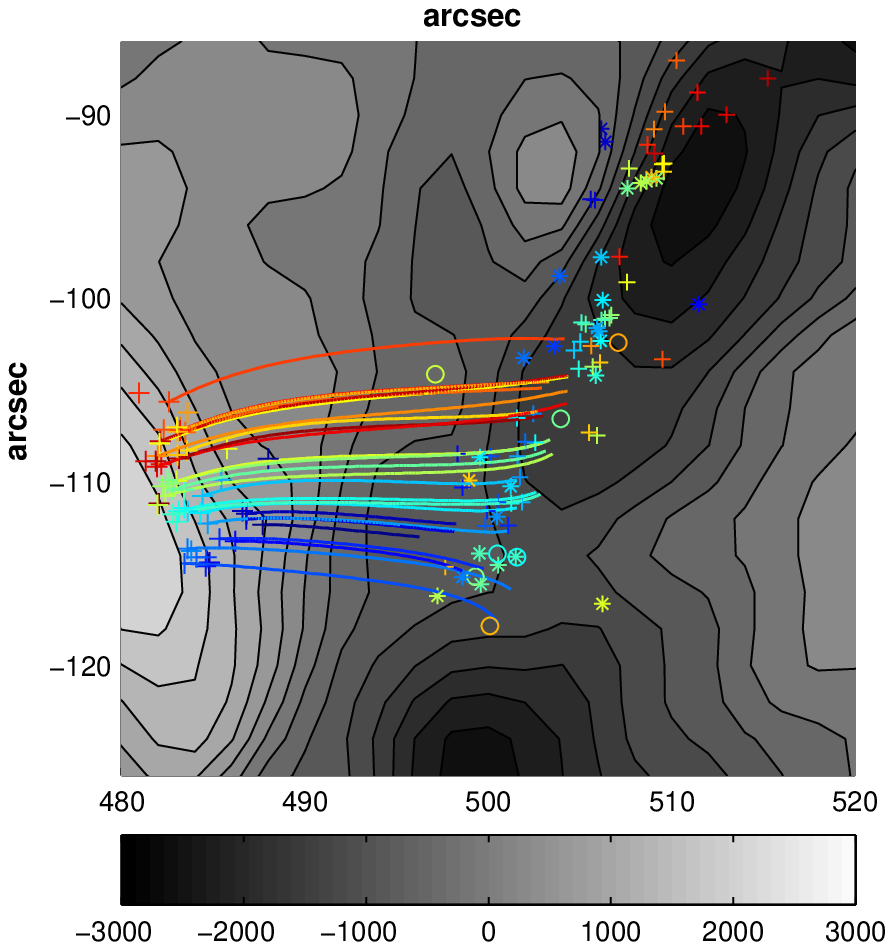}
              }
     \vspace{-0.43\textwidth}   
     \centerline{\Large \bf     
      \hspace{0.08 \textwidth}  \color{white}{(a)}
      \hspace{0.41 \textwidth}  \color{white}{(b)}
         \hfill}
         
     \vspace{0.41\textwidth}
    
\caption{Morphology of the sources of hard X-ray emission and magnetic field in the region of the 29 May 2003 flare. \textbf{(a, b)} Gray-scale contours represent longitudinal magnetic fields at levels $\pm{(0, 300, 600, 900, 1200, 1500, 1800, 2100, 2400, 2700)}$ \textrm{G} obtained with SOHO/MDI (at about 00:02 UT; gray color bar in gauss). \textbf{(a)} Centroids of the brightest RHESSI eastern HXR source at $30 - 100$ \textrm{keV} and at different times are marked by the colored thick crosses (exposure time for the RHESSI image reconstruction is 8 \textrm{s}; colored color bar indicates an appropriate time from 01:01:00 till 01:08:00 UT). Centroids of three the brightest RHESSI western HXR sources at $30 - 100$ \textrm{keV} at each time are marked by thin crosses, asterisks and circles (in descending order of their brightening). Centroids of the RHESSI thermal HXR source at $12 - 18$ \textrm{keV} are marked by dots. The magnetic inversion line is marked by the light green lines. \textbf{(b)} Potential magnetic field lines started from every second RHESSI eastern HXR source and the same RHESSI non-thermal HXR sources as in \textbf{(a)}.     
           }              
   \label{F-5}
   \end{figure}

It is well seen that both QPPs and the spikes after the QPP-phase are emitted from different positions (Figure~\ref{F-5}). Velocity of the eastern source is calculated as a displacement of its centroid position in each time divided by the time between two consecutive images taken for calculation (8 s; Figure~\ref{F-3}(b)). An average velocity is about $127$ \textrm{km s$^{-1}$}. But the source reveals abrupt jumps in some moments, which mainly coincide in time with growth and decay phase of each pulsation. The peaks of the QPPs practically coincide with the slowest velocity of the source (in the range of the calculated errors the source can be practically considered as motionless in these time intervals, except the QPP 1). At the average the source moves more in the direction parallel to the approximated magnetic inversion line, than in the perpendicular one. The average parallel component of the velocity, $v^{\|}_{ftp}$, is about $94$ \textrm{km s$^{-1}$} and the perpendicular component, $v^{\bot}_{ftp}$, is $65$ \textrm{km s$^{-1}$}.     

\begin{figure}    
   \centerline{\includegraphics[width=0.99\textwidth,bb=128 338 461 507,clip=]{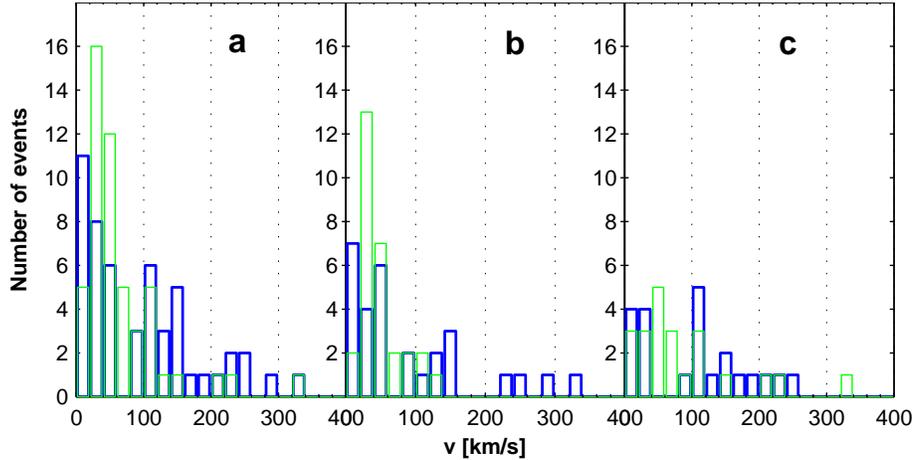}
              }
              \caption{Distribution of the absolute values of the eastern non-thermal HXR source velocity components perpendicular (green) and parallel (blue) to the average magnetic inversion line during the QPP and post-QPP-phase together (01:01--01:08 UT; \textbf{a}); during the QPP-phase only (01:01--01:05 UT; \textbf{b}); during the post-QPP-phase only (01:05--01:08 UT; \textbf{c}) of the 2003 May 29 solar flare. 
                      }   \label{F-4}
   \end{figure}

To examine the standard 2D model more quantitatively we also calculate magnetic field below the eastern source in each time as the average field of the nearest 9 pixels of the SOHO/MDI magnetogram (made at 00:02 UT) below the centroid of the source (Figure~\ref{F-3}(c)). The product $v_{ftp}B_{ftp}^{2}$ has similar time profile to that of the velocity (\cf~Figure~\ref{F-3}(d) and Figure~\ref{F-3}(b)), because of the quite smooth character of the calculated $B_{ftp}$ time profile relatively to the time profile of $v_{ftp}$. Nevertheless, some correlations between the HXRs time profile and $v_{ftp}B_{ftp}^{2}$ can be found in some moments, mainly after the QPP-phase. It is interesting to note that the similar behavior of the RHESSI HXR sources was found by \inlinecite{Krucker05} during the late phase of their flare. This may indicate that in the decay phase of some flares magnetic field becomes less complicated than in the initial phase (maybe, it tends to the non-sheared state), and the simple 2D model can be more pronounced. The next fact is in favour of that. Although, the curves of the absolute values of the both velocity components relatively to the averaged magnetic inversion line are quite similar indicating simultaneous motions in the both directions, absolute values of the parallel component prevail over the perpendicular ones in the range $>120$ \textrm{km s$^{-1}$} during the QPP-phase, but the perpendicular component in this range reveals itself stronger in the post-QPP-phase (Figure~\ref{F-3}(b)). Histograms on Figure~\ref{F-4} show this a bit more clear (unfortunately, we have a small amount of data points). 

Magnetic field is on average weaker in the side of the western source and has more complex topology than in the region of the eastern one. Field lines can diverge harder in the western side of the magnetic inversion line to save magnetic flux, consequently accelerated electrons can precipitate into a wider region here producing a more spacious HXR sources. To roughly examine this we calculated the potential magnetic field inside the flare region using the Green-function method. Figure~\ref{F-5}(b) shows contours of the SOHO/MDI magnetogram (made at about 00:02 UT, prior to the flare) overlaid by the calculated potential field lines originated from centroids of the eastern source (every second centroid is used from the series of 8 s images for clarity). It is seen, that magnetic field lines have a fan-like shape. Thus, more than one non-stationary HXR sources (mainly 2 or 3) can be naturally visible inside the western ribbon in some moments during the flare (Figure~\ref{F-2}(d)), that is really occurred some moments during the flare, whereas the eastern source is mainly single. To all appearance, different western sources can not be resolved and merge into one alongated source in the other moments, due to the limited resolution and dynamic range of the RHESSI observations. No systematic motion of the western HXR sources is found during the QPP-phase of the flare. The sources appear in quite random places near the western ribbon. The sources are clear separated and moves mainly in the north-west direction along the western ribbon after 01:05 UT. This seems to be because of the footpoints of the flaring loops are situated mainly in the strong magnetic field of the north-west sunspot during this time period (Figure~\ref{F-5}), so precipitating electrons can be more strongly collimated giving a clear source of HXRs that can be resolved by RHESSI.

Note, that the most footpoints of the model magnetic field lines do not coincide well with the observed western hard X-ray sources indicating quite large deviation from potentiality or that electrons are injected not in a single magnetic field line inside the reconnection region each moment. Also magnetic field in the active region could change significantly between the time when the magnetogram was made and the time of the flare. And moreover magnetic field could hardly change under the process of magnetic reconnection in the course of the flare.

\section{Flare of 2005 January 19} 
  \label{S-January}

\subsection{Features of the flare} 
  \label{S-JanFeature}  

An eruptive two-ribbon flare (N19\textdegree{}, W47\textdegree{}) of the GOES class X1.3 within the powerful NOAA Active Region 10720 started at 08:03 UT, had its peak at 08:22 UT and finished at 08:40 UT in the GOES soft X-ray range (according to \textit{Solar-Geophysical Data}). The halo coronal mass ejection was firstly seen at 08:29:39 UT by the SOHO/Lasco/C2 but no active filament was observed in H$\alpha$ (according to \textit{Solar-Geophysical Data}). ACS observations of the flare in the HXR range was clogged by solar energetic particles from the X3.8 flare of the 17$^{th}$ January. But, the flare was fully observed by RHESSI. The flare microwave emission was observed by the Learmonth Radio Telescope (Australia) which is a part of the Radio Solar Telescope Network operated by the US Air Force.  

\begin{figure}    
   \centerline{\includegraphics[width=0.99\textwidth,bb=134 282 465 564,clip=]{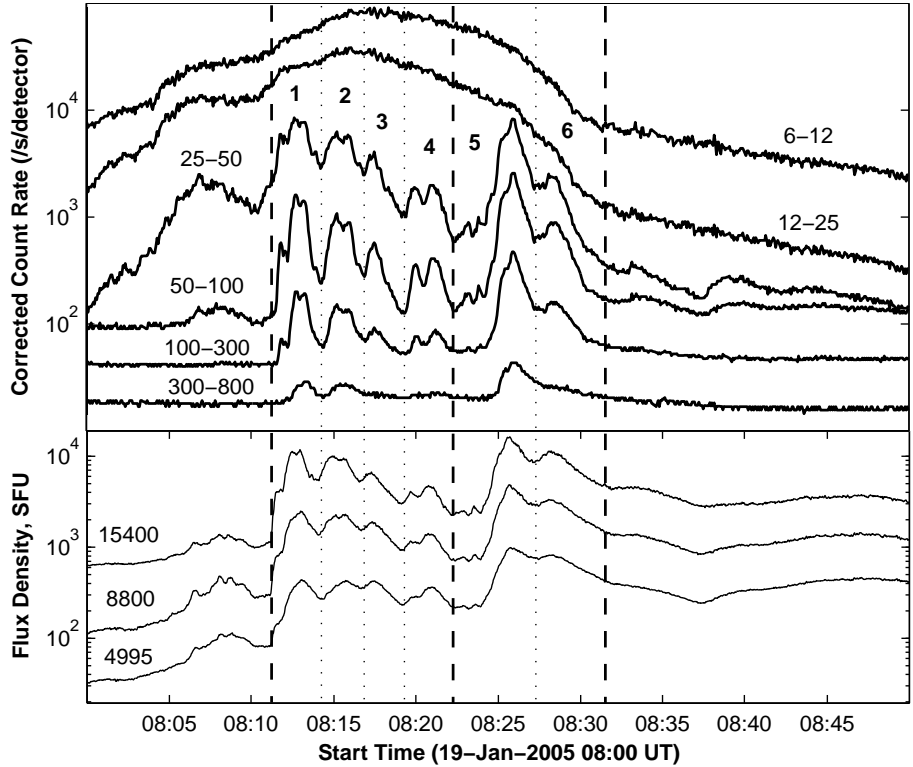}
              }
              \caption{\emph{Top panel:} RHESSI corrected count rates at six energy bands (in \textrm{keV}; counts are multiplied by 3, 1, 2, 1.5, 0.5, 0.3 starting from the lowest band) accumulated over the detectors 1, 3, 4, 5, 6, 9. The time resolution is 4 \textrm{s}. The QPPs are numerated by $1-6$. \emph{Bottom panel:} Flux density of solar microwave emission on three fixed frequencies (flux at 8800 and 4995 \textrm{MHz} are multiplied by 0.3 and 0.1 respectively for clarity) observed by the Radio Telescope in Learmonth. The time resolution is 1 \textrm{s}. The background is not subtracted.     
                      }   \label{F-6}
   \end{figure}

Light curves of the microwave and the full-sun HXR emissions are shown on Figure~\ref{F-6}. Four QPP-structures of the microwave and HXR emission in the $25-300$ \textrm{keV} range with durations of about 3 minutes are clearly seen during 08:12 - 08:23 UT (we will call this time interval ``the distinct QPP-phase''; it is between two left vertical dashed lines). These four structures are clearly composed of several merged shorter spikes. This is similar to that observed during two famous flares on 1980 June 7 and 1982 November 26 (\opencite{Tajima87}, and references therein). Spikes 5 and 6 are separated from the first four ones by the long gap of about 2 minutes which also contains several shorter spikes. All pulsatory structures during 08:12 - 08:32 UT we will call ``the QPP-phase''. Light curves of thermal HXR emission $<25$ \textrm{keV} are more smooth and do not reveal QPPs.

Detailed spectral analysis of the RHESSI HXR emission during the flare were made by \inlinecite{Grigis08} and \inlinecite{Saldanha08}. They found that practically all main spikes of non-thermal HXR emission ($>50$ \textrm{keV}) during the QPP-phase show the soft-hard-soft behavior. The spectral behavior changes to the pronounced soft-hard-harder state about six minutes after the end of the QPP-phase (at about 08:38 UT).

\subsection{Images of the flare region} 
  \label{S-images2}

\begin{figure}    
   \centerline{\hspace*{0.01\textwidth}
               \includegraphics[width=0.498\textwidth,bb=90 26 390 321,clip=]{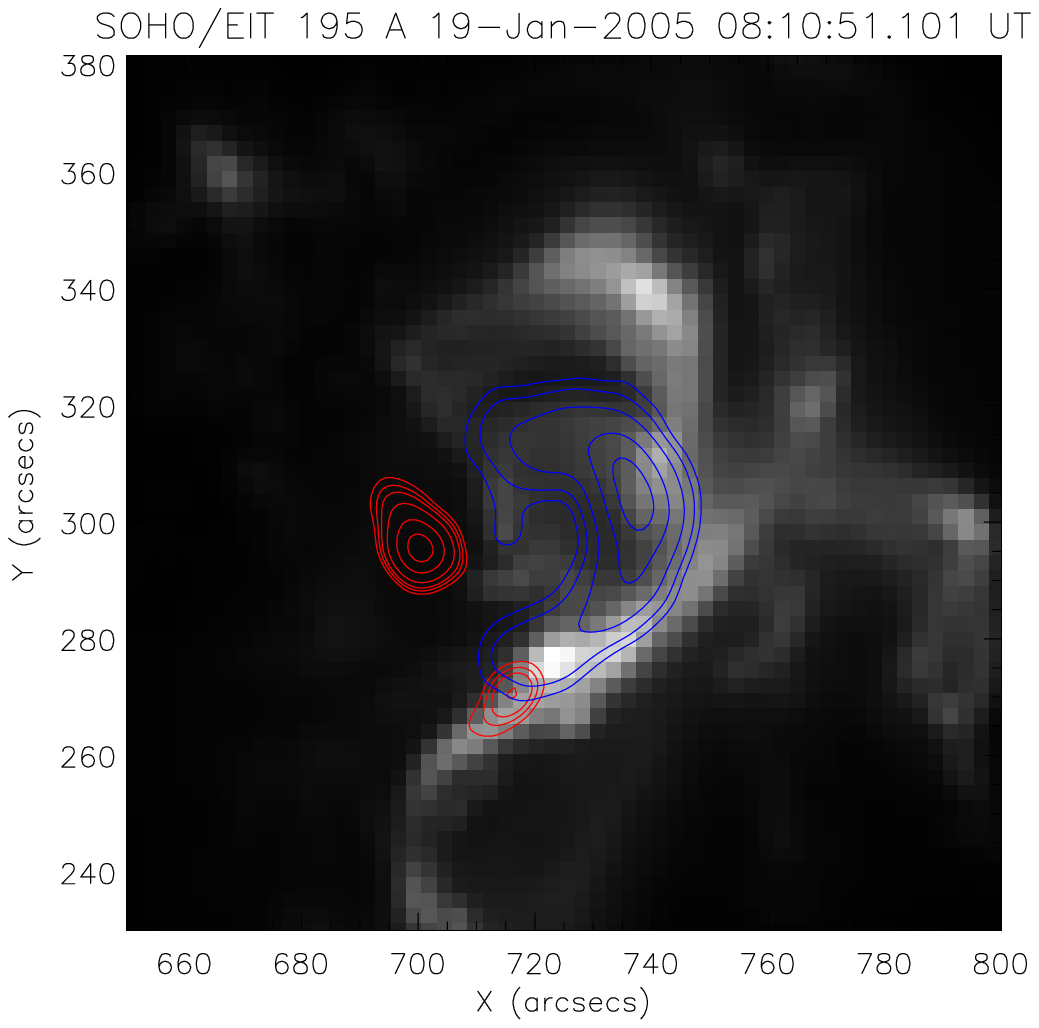}
               \hspace*{-0.02\textwidth}
               \includegraphics[width=0.498\textwidth,bb=90 26 390 321,clip=]{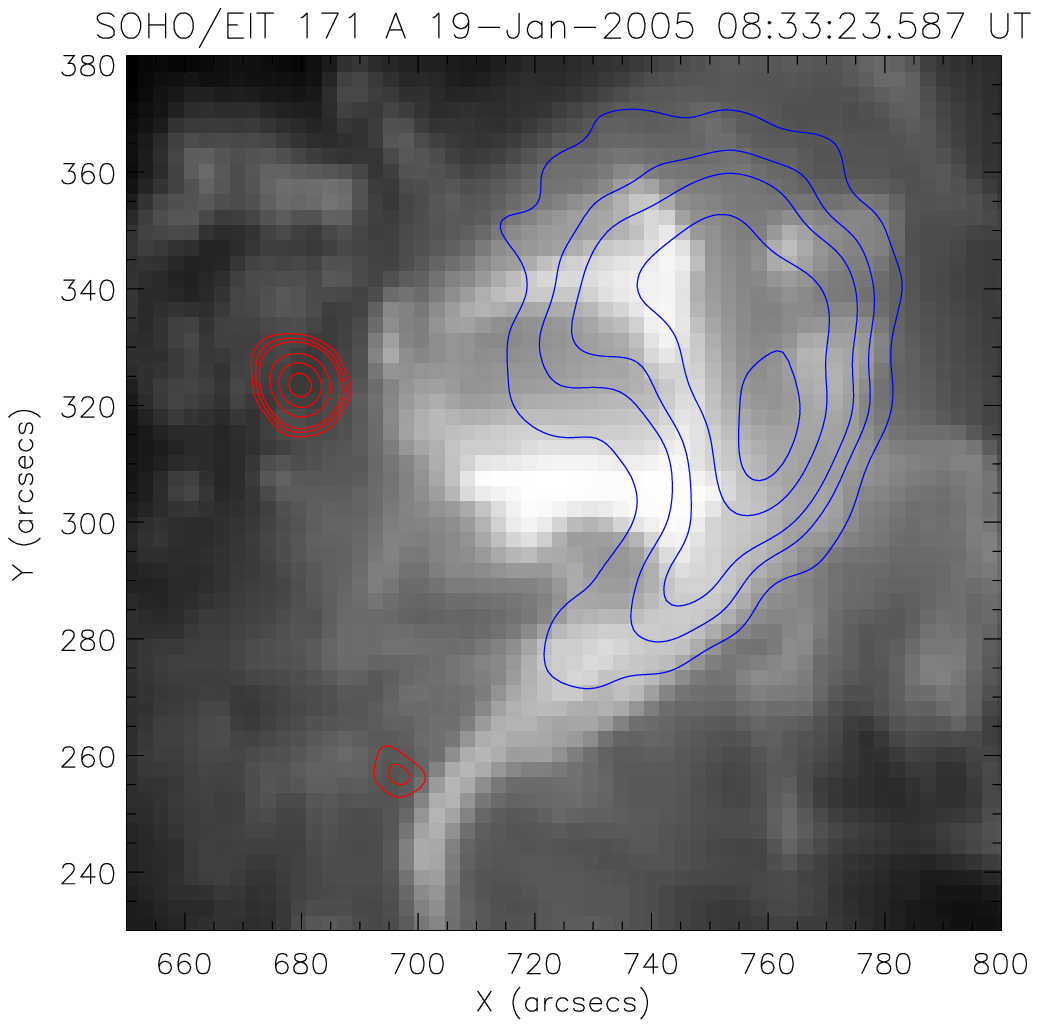}
              }
     \vspace{-0.44\textwidth}   
     \centerline{\Large \bf     
     \hspace{0.08 \textwidth}  \color{white}{(a)}
     \hspace{0.4\textwidth}  \color{white}{(b)}
         \hfill}
         
    \vspace{0.41\textwidth}
     \centerline{
                 \hspace*{0.01\textwidth}
                 \includegraphics[width=0.498\textwidth,bb=90 26 390 321,clip=]{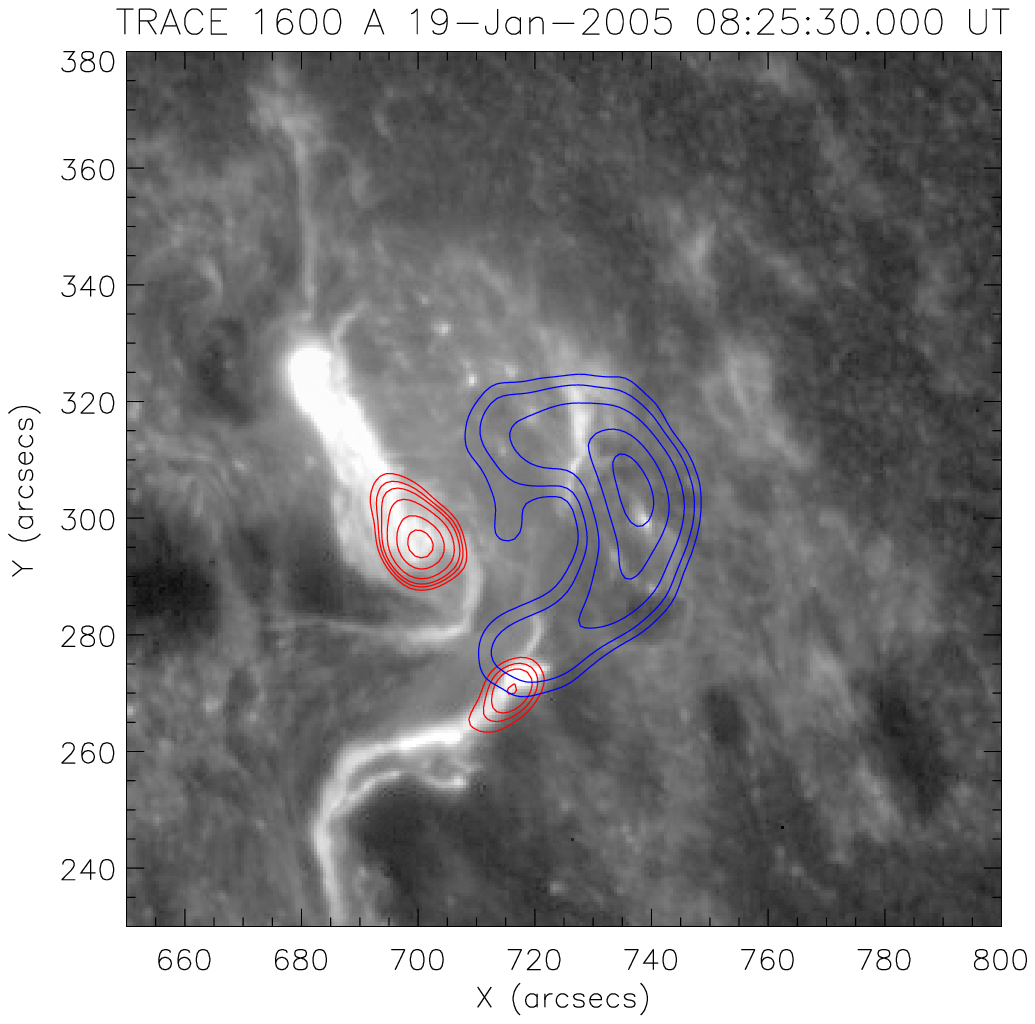}
                 \hspace*{-0.02\textwidth}
                 \includegraphics[width=0.498\textwidth,bb=90 26 390 321,clip=]{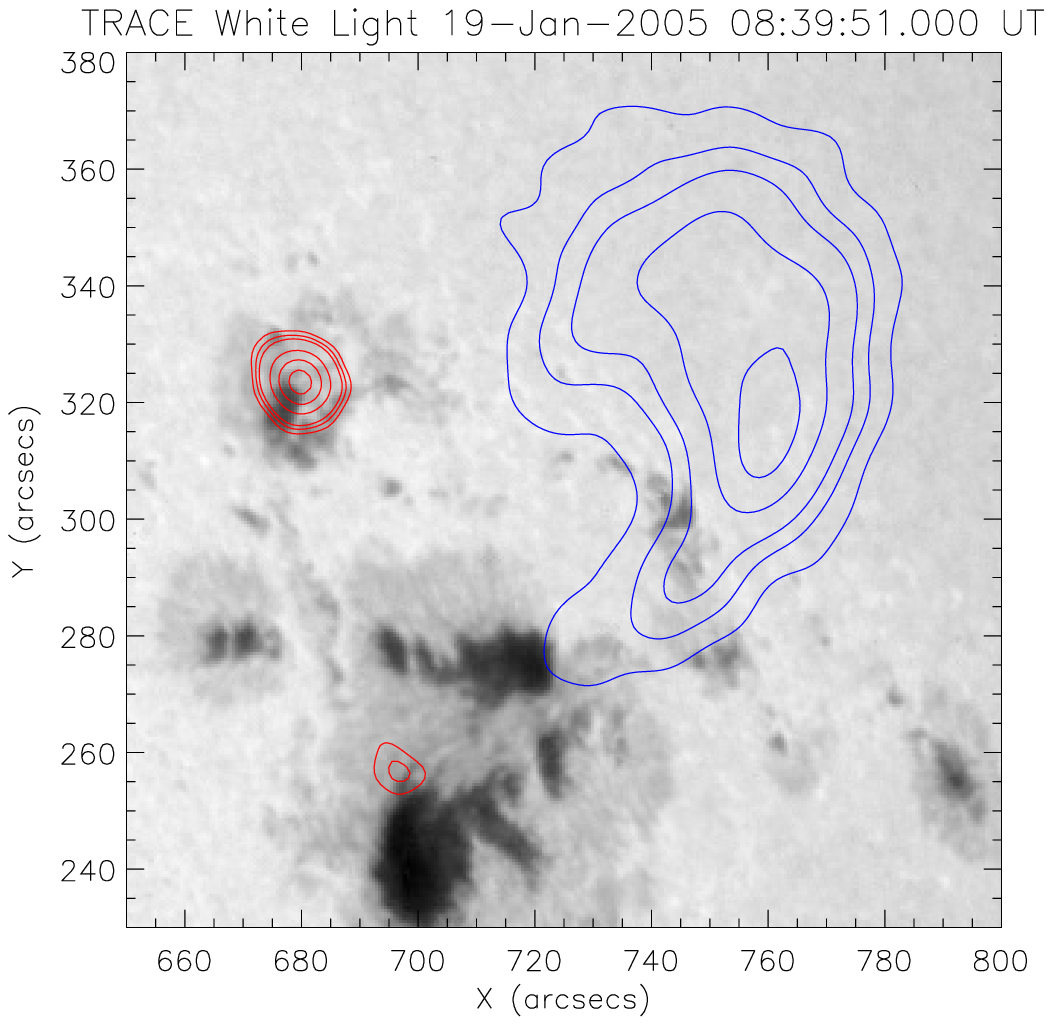}
                 }                   
     \vspace{-0.42\textwidth}
     \centerline{\Large \bf     
      \hspace{0.08 \textwidth}  \color{white}{(c)}
      \hspace{0.4\textwidth}  \color{black}{(d)}
      \hfill}
      \vspace{0.4\textwidth}
\caption{Morphology of emission in different wavelengths in the region of the two-ribbon solar flare on 19 January 2005. \textbf{(a, c)} The SOHO/EIT 195 \textrm{\r{A}} and the TRACE 1600 \textrm{\r{A}} images overlaid by the RHESSI $50-100$ \textrm{keV} contours (30\%, 35\%, 40\%, 50\%, 70\%, 90\%; red lines) made at 08:11:40 - 08:12:40 UT and $12-18$ \textrm{keV} contours (35\%, 40\%, 50\%, 70\%, 90\%; blue lines) made at 08:13:48 - 08:15:56 UT. These images were made during the QPP-phase and were rotated to 08:15:00 UT. \textbf{(b, d)} The SOHO/EIT 171 \textrm{\r{A}} and the TRACE white light images overlaid by the RHESSI $50-100$ contours (20\%, 25\%, 30\%, 50\%, 70\%, 90\%; red lines) and $12-18$ \textrm{keV} contours (30\%, 40\%, 50\%, 70\%, 90\%; blue lines) made at 08:40:00 - 08:43:00 UT, after the QPP-phase. The images were rotated to 08:40:00 UT.
                      }              
   \label{F-7}
   \end{figure}

Figure~\ref{F-7} shows typical morphology of the neutral emission in the flare region. Two sources of non-thermal HXR emission in $50 - 100$ \textrm{keV} range place near the footpoints of the loop arcade which is seen in the 171 and 195 \textrm{\r{A}} images made by the EUV Imaging Telescope onboard SOHO satellite (SOHO/EIT; Figure~\ref{F-7}(a, b)). These sites practically coincide with the conjugated TRACE 1600 \textrm{\r{A}} ribbons (\cf~Figure~\ref{F-7}(c)). The northern HXR source is systematically much brighter than the southern one, which is placed mainly between two sunspots (Figure~\ref{F-7}(d)). Thermal HXR source in $12 - 18$ \textrm{keV} range places mainly inside the region of the bright EUV loops.

To trace evolution of the HXR sources at 08:11:50 - 08:50:00 UT (this time interval includes the QPP-phase and some time after it) we reconstruct sequences of the RHESSI $12 - 18$ \textrm{keV} (thermal source) images accumulated over 120 \textrm{s} (about 30 periods of the RHESSI rotation) and of $50 - 100$ \textrm{keV} (non-thermal sources) images accumulated over 20 and 60 \textrm{s}. Clean and Pixon algorithms were applied to the RHESSI data from detectors $2 - 8$. The best image was chosen each time. Longer times of photons accumulation are required to define a centroid position of the thermal and the dull southern non-thermal HXR source accurately (at the same time we lose some information on their dynamics).  

\begin{figure}    
   \centerline{\includegraphics[width=0.5\textwidth,bb=162 292 412 540,clip=]{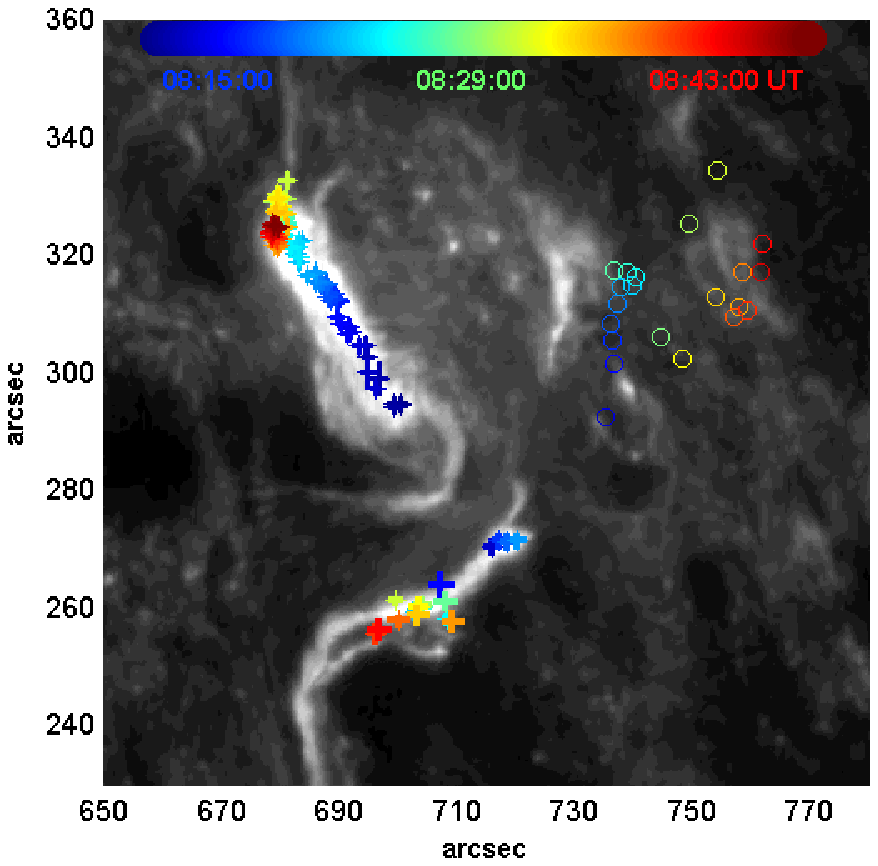}
              }
              \caption{Dynamics of the main HXR sources at the time of QPP-phase and after it during the solar flare on 2005 January 19 (08:11:50 - 08:50:00 UT). The centroids of the HXR sources are superposed on the TRACE 1600 \textrm{\r{A}} image (made at 08:25:30 UT), which indicates two main flare ribbons. The centroids position of the non-thermal $50 - 100$ \textrm{keV} HXR sources are marked by crosses. The size of each cross indicates an error of the centroid position calculation. The centroids of the northern and southern sources are taken every 20 and 60 \textrm{s} respectively, because of a dullness of the southern source. The centroid of the thermal ($12 - 18$) \textrm{keV} source is obtained every 120 \textrm{s} and marked by circles.
                      }   
                      \label{F-8}
   \end{figure}

Figure~\ref{F-8} shows the TRACE 1600 \textrm{\r{A}} image of the flare ribbons (the first 1600 \textrm{\r{A}} image for this flare; made at 08:25:30 UT) overlaid by the reconstructed centroids of the RHESSI $50-100$ \textrm{keV} HXR sources at 08:11:50 - 08:50:00 UT. The southern non-thermal HXR source is quite dull and composed of 2 or 3 separated sources in some moments. So, it is difficult to trace its motion in details, although it moves generally to the south-east. The thermal HXR source moves mainly to the north-west in the reconstructed 2D images and, apparently, somehow upward in reality. But it is composed of several equivalent sources in some moments and it is also difficult to trace the sources dynamics in details.

\begin{figure}    
   \centerline{
               \includegraphics[width=0.98\textwidth,bb=130 260 451 571,clip=]{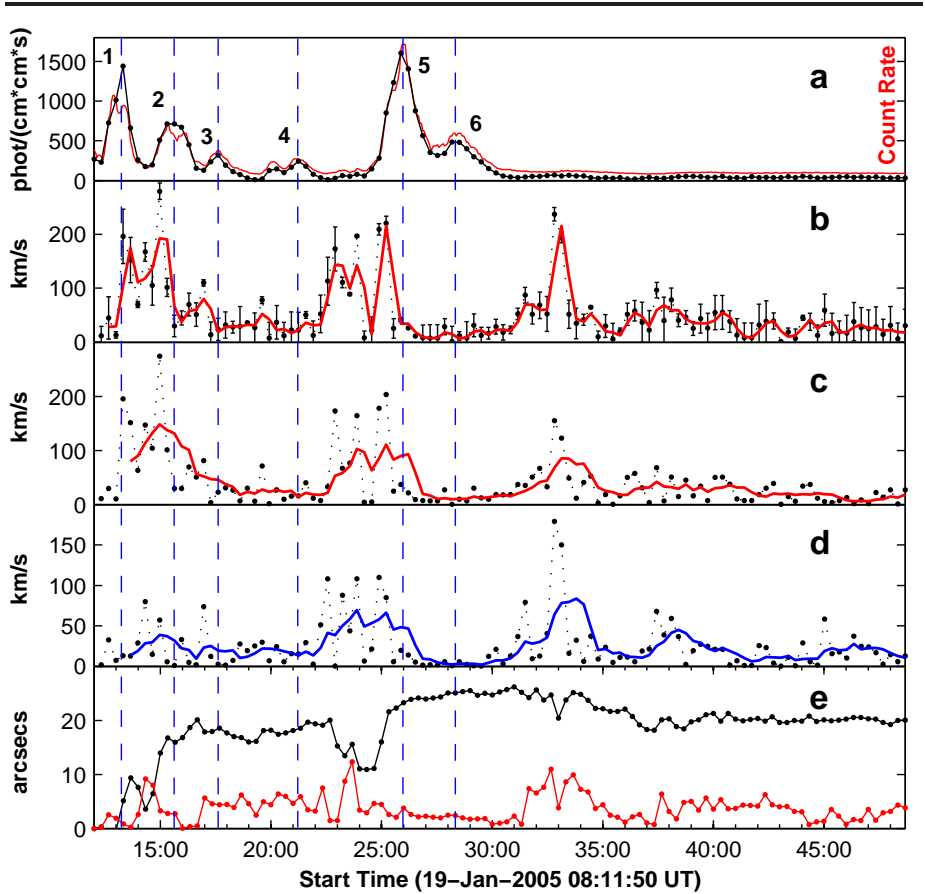}
              }
              \caption{Temporal variations of the northern non-thermal $50 - 100$ \textrm{keV} HXR source during the flare on 2005 January 19. \textbf{(a)} RHESSI full-sun corrected count rate at $50 - 100$ \textrm{keV} accumulated over the detectors 1, 3, 4, 5, 8, 9 each 4 \textrm{s} (red line) and the flux of $50 - 100$ \textrm{keV} HXRs calculated from the 20-s Pixon images of the northern source (black dotted line; the flux was multiplied by 40). \textbf{(b)} Full velocity of the northern source (black dotted line). Red line indicates the time profile of the source velocity smoothed over 40 \textrm{s}. Black vertical dashes with dots indicate an error of the velocity calculation. \textbf{(c, d)} Parallel and perpendicular components of the northern source velocity relatively to the approximated direction of the northern flare ribbon calculated each 40 s (black dots) and smoothed over 100 \textrm{s} (red and blue lines respectively). \textbf{(e)} The total displacement of the northern source along (black dotted line) and perpendicular (red dotted line) with respect to the northern ribbon vector. The pulsations are numerated according to Figure~\ref{F-6} and their peaks are marked by blue vertical dashed lines. 
                      }   
                      \label{F-9}
   \end{figure}

Thus, we concentrate on the detailed investigation of the brightest northern non-thermal HXR source, which moves systematically to the north-east till about 08:31 UT (Figure~\ref{F-8} and Figure~\ref{F-9}(e)). The path traveled by the source is coincide well with the northern flare ribbon, which can be roughly approximated by the vector which is directed to the north-east. This may indicate that the increased UV emission from the entire ribbon (observed firstly about 15 minutes after the start of non-thermal HXR emission) can be caused by sequentially precipitating electrons in adjacent arcade loops (\opencite{Fletcher01}). Unfortunately, there were not TRACE observations of the flare region during these first 15 minutes to trace dynamics of the ribbon development. The main flux of non-thermal HXR emission comes from this northern source. Temporal variations of the flux from this source coincide well with the RHESSI full-sun count rate in $50 - 100$ \textrm{keV} range (Figure~\ref{F-9}(a)).

\begin{figure}    
   \centerline{\includegraphics[width=0.98\textwidth,bb=128 341 467 502,clip=]{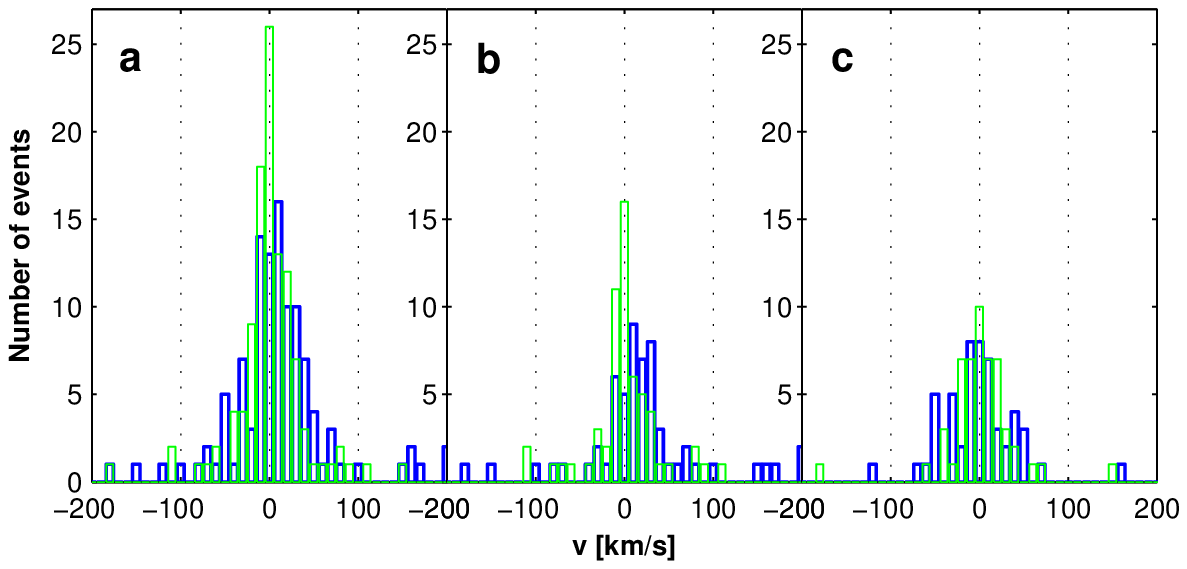}
              }
              \caption{Distribution of the northern non-thermal HXR source velocity components perpendicular (green) and parallel (blue) to the approximating vector of the TRACE 1600 \textrm{\r{A}} northern flare ribbon during the QPP and post-QPP-phase together (08:11:50--08:50:00 UT; \textbf{a}); during the QPP-phase only (08:11:50--08:32:00 UT; \textbf{b}); during the post-QPP-phase only (08:32:00--08:50:00 UT; \textbf{c}) of the 2005 January 19 solar flare. 
                      }   \label{F-10}
   \end{figure}

Velocity of the northern source is calculated as in Section~(\ref{S-images1}). The average velocity is about $59$ \textrm{km s$^{-1}$} during the QPP-phase. It is seen that during the QPP-phase there are some bursty increases of the source velocity mainly along the ribbon vector (Figure~\ref{F-9}(c, d)), although there is not a peak-to-peak coincidence between the velocity profile and that of the flux. Moreover, there is a sharp increase of the source velocity after the QPP-phase (about 08:31 - 08:35 UT) without a comparable signature in the curve of the HXR emission, although small increases are seen (Figure~\ref{F-6}). Components of the total source displacement relatively to the ribbon vector indicate the source motions mainly along the general ribbon direction during the QPP-phase, some backward movements during 08:31 - 08:39 UT and practically immobility after this time. This is partially seen from the histograms on Figure~\ref{F-10}. Note, that magnetic field data is not used to examine the rate of magnetic energy release via the relation~(\ref{Eq_2}), because the flare site is at a large distance from the solar disk center.                

     \section{Summary and Discussion} 
  \label{S-Discussion}
Analysis of the RHESSI observations of two solar flares which were accompanied by minute QPPs of non-thermal HXR and microwave emission reveals motions of thermal and non-thermal HXR sources. Microwaves and the thermal HXR emission were emitted mainly from the arcade apex, but non-thermal HXRs were emitted successively from the footpoints of the adjacent coronal loops stacked into the arcades. This is especially seen for the flare on January 19, when the HXR source was systematically moving along the flare ribbons about 20 minutes. The general velocity component of the moving sources was directed parallel to the magnetic inversion line (the May 29 flare) or parallel to the flare ribbon (the January 19 flare). The peak-to-peak correlation between the velocity of the HXR sources and the flux of non-thermal emission was not found. Moreover, anticorrelation between the mentioned parameters is observed during some moments of the flare. The sources movement looks like abrupt jumps mainly along the arcade and sitting a bit during the QPPs' maxima. These may indicate a non-validity of the simple 2D model of the eruptive flares based on the magnetic reconnection process in the observed events. What useful information could be obtained from these imaging observations to understand the observed quasi-periodicity of non-thermal emission and its relation to the processes of primary energy release? 

First, we can conclude, that the separate acts of energy conversion from accelerated electrons into the neutral emission have occurred by piecemeal in different flaring loops rather than in a single one (we note that as before we can't observe the sites of primary energy release). Thus, that models of the flare QPPs, based on a single oscillatory flaring loop, don't work here. Apparently, this is the only evident conclusion from our observations. The following is just a discussion.

Probably, an interacting loop model proposed by \inlinecite{Emslie81} also does not work in its direct form in the observed events. According to this model the separation distance between two interacting and bursting loops, $D$, can be roughly estimated as 
\begin{equation}  \label{Eq_3}
     D \sim v_{A}T\approx\frac{2.18\times{10^{11}}\times{B}\times{T}}{\sqrt{n}}
   \end{equation}
in cm, where $v_{A}$ is the Alfven velocity, $B$ is magnetic field in \textrm{G}, $T$ is the time between two consecutive bursts in \textrm{s}, $n$ is the electron density in \textrm{cm$^{-3}$}. In the case of the May 29 flare $T\approx{60}$ \textrm{s} and $B\approx{500}$ \textrm{G}  near the arcade apex as it was estimated from the calculated potential field. The emission measure is $\approx(1.6 - 5.4)\times{10^{49}}$ \textrm{cm$^{-3}$} from the GOES observations during the QPP-phase. One can roughly interpret the emitting volume, $V$, as the entire flare arcade which can be approximated by a half of cylinder with the length of $\approx40^{\prime\prime}$ or $\approx3\times10^9$ \textrm{cm} and the diameter of $\approx25^{\prime\prime}$ or $\approx1.8\times10^9$ \textrm{cm}. Hence, $V\approx3.8\times10^{27}$ \textrm{cm$^3$} and $n\approx{(0.7 - 1.2)\times10^{11}}$ \textrm{cm$^{-3}$}. Thus, $D\approx{(1.9-2.5)\times{10^{10}}}$ \textrm{cm} that is one order of magnitude larger than the length of the entire flare arcade. Even if we assume $B\approx{50}$ \textrm{G}, $D$ will be equal to the length of the arcade. But $D$ must be less than the arcade length, because observations show that non-thermal HXRs are emitted from the arcade foots. The same we can expect for the January 19 flare, which has $T\approx{180}$ \textrm{s} and $n\approx{(0.4-0.6)\times10^{11}}$ \textrm{cm$^{-3}$}. Unfortunately, we don't have information about magnetic field, but it is reasonably to suppose $B\approx100$ \textrm{G}, that is less than it was supposed for the May 29 flare, because the sunspots are also less in this active region. Thus, $D\approx{(1.6-2)\times10^{10}}$ \textrm{cm}, that is also a very large value. Therewith, the evaluated Alfven velocity is $\sim{1000}$ \textrm{km s$^{-1}$}, that is several times larger than the observed velocity of HXR sources (the same was observed by \opencite{Grigis05}). From here we can also conclude that a magnetosonic wave with the velocity $V \approx{\sqrt{v_{A}^{2}+v_{S}^{2}}}$, where $v_{A}<<c$ and $v_{S}$ is the sound velocity, can not also be considered as the propagating trigger of the primary energy release as it was suggested by \inlinecite{Vorpahl76}.
As far as the standard 2D reconnection model of the flare does not work properly in the observed events and the potential magnetic field lines don't conjugate the observed HXR sources well, one can expect that the observed arcades are sheared. So, different modes of the MHD waves may be presented, but it is beyond the scope of this paper to discuss them.    

A scenario based on an oscillatory loop arcade as a whole, which is sometimes observed (\eg{}, \inlinecite{Verwichte04} have observed standing kink oscillations of post-flare arcade loops), can not be ruled out with certainty, although there were not direct observations of such loops behavior in the examined events. We can imagine two principally different possibilities of such models to be without going into details in this work:\\
\textbf{(a)} consecutive involvement of similar and adjacent arcade loops into the processes of primary energy release, acceleration, and precipitation of electrons, like in an interacting loop model but via, say, transverse oscillations in arcade loops and further via the mechanism proposed by \inlinecite{Nakariakov06};\\
\textbf{(b)} simultaneous trapping of the bulk of accelerated electrons by an entire flaring arcade and consecutive pulsed precipitation in different loops, because of that sort of MHD waves which can propagate along the arcade and decrease the magnetic mirror ratio in each loops. The \textbf{(b)}-type models seem quite doubtful for the explanation of the observed events. It is difficult to imagine a long effective trapping of many electrons without a significant decrease of their amount as it was in the case of the QPPs 5 and 6 of the January 19 flare about $15-20$ minutes after the first non-thermal spike. Therewith, imaging analysis of the X-class flare on 2005 January 17 from the same active region with the 19$^{th}$ January flare made by \inlinecite{Temmer07} in H$\alpha$ and HXR ranges shows that the bulk of electrons was mainly accelerated and injected into a certain set of arcade loops and only a small fraction of electrons was injected to the entire flare arcade. Though, ribbon-like HXR sources were observed, although it is a rare phenomenon (\inlinecite{Masuda01}; \inlinecite{Jing07}).

Evidence in favor of the \textbf{(a)}-type scenario can be found in the work of \inlinecite{Minoshima08}. Authors compared the observed behaviors of the HXR and microwave emission with their numerical simulations of the static ``trap-plus-precipitation'' model (\eg{}, \opencite{Melrose76}) for two adjacent pulsations of the May 29 flare. In the frame of their model the observed behaviors of the separate pulsations can be qualitatively explained by the separate injections of accelerated electrons into a magnetic loop having the observed parameters. Authors considered the same loop parameters for the both spikes of emission. Such approach could occur even in the case of the separate traps, because the loops are quite similar in the concerned part of the arcade (Figure~\ref{F-5}). Also, the soft-hard-soft spectral behavior of the HXR emission was successfully explained by an electron injection function with the same spectral property. 

For all that, from our observations it is impossible to finally conclude what physical mechanism forces primary energy to be released quasi-periodically in different places of the active region. We have to wait new simultaneous imaging observations in the microwave, EUV, and HXR ranges of solar flares with quasi-periodic pulsatory non-thermal emission.
           
\clearpage{}

\end{article} 
\end{document}